\documentclass[twocolumn,showpacs,preprintnumbers,amsmath,amssymb]{revtex4}

\usepackage{graphicx}
\usepackage{dcolumn}
\usepackage{bm}

\begin{document}

\title{Injection of a cold atomic beam into a magnetic guide}

\author{C. F. Roos, P. Cren, T. Lahaye, J. Dalibard and D. Gu\'ery-Odelin}

\address{Laboratoire Kastler Brossel$^*$, Ecole Normale Sup\'erieure \\
24, Rue Lhomond, F-75231 Paris Cedex 05, France}

\date{\today}

\begin{abstract}
We report the continuous or pulsed loading of a slow and cold atomic beam into a magnetic guide. In
order to optimize the transfer into the guide, we have studied two coupling schemes. The first one is
based on an auxiliary two-dimensional MOT that confines the atoms transversally before they enter the
guide. However, for low atomic velocities, the atoms are strongly heated in the longitudinal
direction. This limitation does not occur in the second coupling scheme which relies on magnetic
confinement in the transfer zone. For this purpose, we have constructed a miniature magnetic guide
located between the atomic source and the long magnetic guide. The latter scheme allows to inject
atoms with velocities down to 70 cm/s into the guide.
\end{abstract}


\maketitle

A spectacular challenge in the  field of Bose-Einstein condensation consists in the achievement of a
continuous beam operating in the quantum degenerate regime. This would be the matter wave equivalent
of a cw monochromatic laser and it would allow for unprecedented performances in terms of
focalization or collimation. In \cite{ScienceK02}, a continuous source of  Bose-Einstein condensed
atoms was obtained by periodically replenishing a condensate held in an optical dipole trap with new
condensates. This kind of technique raises the possibility of realizing a continuous atom laser.  An
alternative way to achieve this goal has been studied theoretically in \cite{Mandonnet00}.
 A non-degenerate, but already slow and cold beam of particles, is injected into a magnetic guide
\cite{Schmiedmayer95,Denschlag99,Goepfert99,Key00,Dekker00,Teo01,Sauer01,Hinds99} where transverse
evaporation takes place. If the elastic collision rate is large enough, efficient evaporative cooling
 can lead to quantum degeneracy  at the exit of the guide.  The condition for reaching
degeneracy with this scheme can be formulated by means of three parameters: the length $\ell$ of the
magnetic guide on which evaporative cooling is performed, the collision rate $\gamma$ at the
beginning of the evaporation stage, and the mean velocity $\bar{v}$ of the beam of atoms. Following
the analysis given in \cite{Mandonnet00}, one obtains
\begin{equation}
\frac{\gamma\ell}{\bar{v}}\gtrsim 500 \; . \label{cond1}
\end{equation}
 Physically, (\ref{cond1}) means that each atom has to undergo
at least 500 elastic collisions during its propagation through the
magnetic guide.

Our experiment aims at implementing this scheme for a beam of $^{87}$Rb atoms. Its success relies
therefore upon two preliminary and separate accomplishments. First, one has to build a source of cold
atoms as intense as possible, with the lowest possible mean velocity. Second, one has to inject the
atomic beam produced by this source into a long magnetic guide with minimal transverse and
longitudinal heating. In the first section of this paper, we describe briefly the way we generate a
high flux of atoms by means of a moving molasses \cite{mm} combined with transverse confinement. This
configuration will be called the {\sl injecting MOT} in the following. In the second section of the
paper, we present the main features of our magnetic guide. In the third section, we focus on how to
efficiently transfer atoms from the injector to the guide, and we report on the performances of our
current experimental setup.

\section{ The injecting MOT}
\label{sec:injector}

\begin{figure}
\begin{center}
\includegraphics[width=8cm]{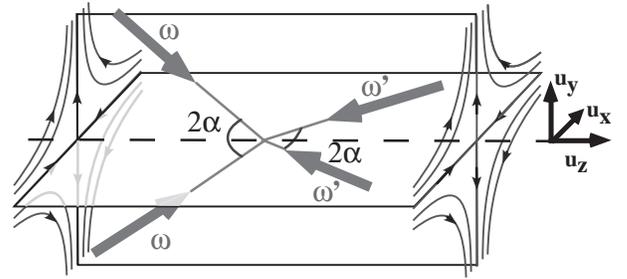}
\caption{\label{MOT}
 Laser and magnetic configurations of the injecting MOT.}
\end{center}
\end{figure}

The injecting MOT  has been described in detail elsewhere \cite{Cren} and we only recall its main
features. It is based upon a four-beam laser configuration similar to the one used for the study of
optical lattices described in \cite{Grynberg93}, superimposed with a magnetic two-dimensional
quadrupole field (see fig.~\ref{MOT}).

The field vanishes along the $z$ axis and the transverse gradient is typically $b'=0.1$~T/m. The
optical arrangement consists of four laser beams in a tetrahedral configuration.  Each beam has a
power of 25 mW and a waist of 12 mm. Two laser beams with frequency $\omega$ propagate in the $yz$
plane along the directions $(\cos\alpha \;{\bf u}_z \pm \sin \alpha\; {\bf u}_y)$ with a positive
helicity. The two other beams with frequency $\omega'$ propagate in the $xz$ plane along the
directions $(-\cos\alpha\; {\bf u}_z \pm \sin \alpha \;{\bf u}_x)$ with a negative helicity. The four
beams are red-detuned with respect to the atomic transition $|5S_{1/2},F=2\rangle\rightarrow
|5P_{3/2},F=3\rangle$, whose frequency is called $\omega_a$. The average detuning $\delta=\bar \omega
-\omega_a$, with $\bar \omega=(\omega+\omega')/2$, is typically $-3\;\Gamma$, where
$\Gamma=2\pi\times 5.9$~MHz denotes the natural width of the excited level of the transition. By
properly choosing $(\omega-\omega')$, one can adjust the mean velocity $\bar v$ from 0 to $\sim
3$~m/s.

The injecting MOT is set up in a rectangular glass cell (130~mm $\times$ 50~mm $\times$ 50~mm).  For
most of the experimental results presented here, the atoms are captured from the low-pressure
background gas (setup A). The partial pressure $P_{87}$ for $^{87}$Rb is measured by the absorption
of a resonant beam. It can be varied from $10^{-9}$ mbar up to the saturated vapor pressure at room
temperature ($3\times 10^{-7}$ mbar) by controlling the temperature of the Rb-reservoir or the
aperture of the intermediate valve. We find that the steady state flux $\Phi$ is independent of the
average velocity of the beam. This flux is proportional to $P_{87}$ in the range $10^{-9}$~mbar $<
P_{87} < 10^{-8}$~mbar, and we obtain:
\[
\Phi= 10^9 \mbox{ atoms/s for } P_{87}=10^{-8} \mbox{ mbar.}
\]

\begin{figure}
\begin{center}
\includegraphics[width=8cm]{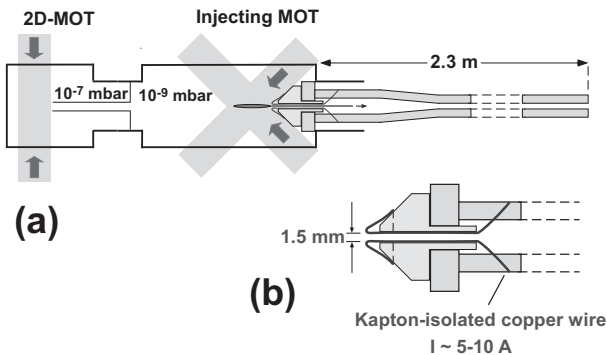} \caption{\label{setup}
 (a) Schematic drawing of the experimental setup. (b) Miniature magnetic guide placed at the
 entrance of the magnetic guide.}
\end{center}
\end{figure}

We have just completed the construction of a new apparatus (setup B, see fig.~\ref{setup}(a)) which
has some advantages over this original system. This apparatus is  based on the same principle as
setup A, but it uses two separate magneto-optical traps. The first one is a two-dimensional
magneto-optical trap (MOT) \cite{Dieckmann98,Pfau02} located in a chamber which has a relatively high
pressure of rubidium vapor ($P \geq 10^{-7}$ mbar). The second MOT (the injecting one) is in a
differentially pumped chamber where the pressure is very low ($P \leq 10^{-9}$ mbar). The first MOT
generates a beam of cold atoms with an average velocity $\sim 40\;$m/s; a  significant fraction of
this beam is captured by the injecting MOT and is subsequently slowed down and cooled. This allows us
to limit the loss of atoms caused by collisions with background gas atoms while they propagate from
the output of the injecting MOT to the entrance of the magnetic guide.

\section{ The magnetic guide}
\begin{figure}
\begin{center}
\includegraphics[width=6.5cm]{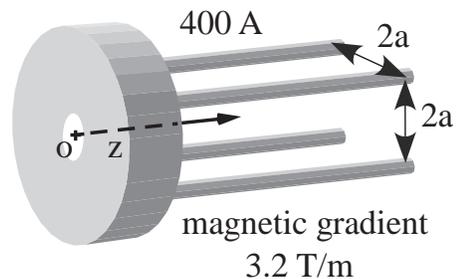}
\caption{\label{guide}
 Schematic drawing of the entrance of the magnetic guide. The hollow metal cylinder close
 to $z=0$ allows for the connection of the electrical currents and the water cooling circuit
 circulating into the four copper tubes.}
\end{center}
\end{figure}

The magnetic guide has a total length of 230~cm. It is made out of four copper tubes (\O$_{\rm
ext}=6$~mm and \O$_{\rm int}=4$~mm) placed in a quadru\-pole configuration (see fig.~\ref{guide}).
The tubes are located in the domain $z>0$ and they are joined in $z=0$ by a hollow metal cylinder,
which allows for the circulation of current and cooling water from tube to tube. The axes of the
copper tubes are placed at coordinates $x=\pm\, a; y=\pm\, a$, with $a=7\;$~mm. A current $I=400$~A
is sent through the tubes, which provides, far from the entrance of the guide (\emph{i.e.} $z \gg
a$), a magnetic gradient $b'=3.2$~T/m in the $xy$ plane.   The transverse magnetic gradient at the
edge of the magnetic guide decreases quite fast  (typical length scale $\sim a$), so it does not
affect the operation of the injecting MOT.

A narrow glass tube ensures differential pumping between the cell of the injecting MOT and the
chamber of the magnetic guide, where the residual pressure has to be minimized to avoid losses due to
collisions with the background gas. We estimate that the pressure in this chamber is $\sim 4\times
10^{-10}$~mbar, which corresponds to a lifetime $\sim 20$~s for the magnetically guided atoms.

\section{ Transfer from the injecting MOT to the magnetic guide}

 We now present the various schemes that we have investigated
to transfer the cold atomic beam generated by the injecting MOT into the magnetic guide. Eq.
(\ref{cond1}) shows that we need a guided beam as intense and cold as possible (a large collision
rate $\gamma$) and at the same time a very low mean velocity $\bar v$.
 We note that, due to the presence of the metal cylinder
located at the entrance of the magnetic guide (see fig.~\ref{guide}), the shortest distance between
the output of the injecting MOT and the entrance of the magnetic guide is $D\sim 25$~mm. In other
words, the injecting MOT is located in the region $-45$~mm $<z<-25$~mm and the atoms have to travel
over a distance of $25$~mm before being captured by the magnetic guide provided that they are in a
magnetic sublevel corresponding to a low-field seeking state.

\subsection{ Free-flight transfer}

 The simplest way to load atoms from the injecting MOT into the
guide consists in letting the atoms propagate freely between the two.  In this purpose, we place a
laser tuned to the $|5S_{1/2},F=2\rangle\rightarrow |5P_{3/2},F=2\rangle$ transition in the region
$-25$~mm~$<z<0$. The repumping laser is blocked in this transfer region, so that the extra
``depumping" laser beam optically pumps the atoms in the $F=1$ ground state. Therefore, provided that
the three magnetic levels $m=0,\pm 1$ of the $|5S_{1/2},F=1\rangle$ are equally populated, we expect
that one third of the atoms (state $m=-1$) emerging from the injecting MOT can be captured by the
magnetic guide.

\subsubsection{ Constraints for free-flight transfer}

 Due to our experimental geometry, there are two issues that
limit the use of a free flight transfer:
\begin{itemize}
\item
It is restricted to relatively large velocities $\bar v > 1.5$~m/s. Otherwise the free-fall due to
gravity, $gD^2/(2\bar v^2)$, is so large that the coupling to the guide becomes velocity-dependent.

\item
During the free flight, the atom jet spreads transversally, which leads to a strong increase of the
transverse temperature $T_\bot$ in the guide, as compared with the temperature in the injecting MOT.
Consider for instance a beam emerging from the injecting MOT with $\bar v=2$~m/s and
$T_\bot=100\;\mu$K. At the entrance of the magnetic guide, the radius of the beam is $R \sim (D/\bar
v)\,\sqrt{k_BT_\bot/M}\sim 1.2$~mm ($M$ is the atomic mass). The corresponding magnetic energy is
\begin{equation}
E_{\rm mag}\sim \mu b'R\sim (D\mu b'/\bar v)\,\sqrt{k_B T_\bot/M}\ . \label{temperature}
\end{equation}
This gives  $E_{\rm mag}/k_B\sim 1$~mK in the present case.

\end{itemize}

\subsubsection{ Pulsed versus continuous mode}

The operation of the injecting MOT in a continuous mode imposes to search for a compromise between
two different requirements. For an efficient loading one has to take a relatively small detuning
($\delta \sim -3\,\Gamma$), so that the trapping force is large. On the contrary, in order to
minimize the temperature of the outgoing atomic beam, a much larger detuning $(\delta \sim-7\,\Gamma$
for our laser intensity) is preferable.

A pulsed operation of the injecting MOT may provide, for a given application, an output beam with
better characteristics than this compromise. First, one loads the injecting MOT with the detuning
which maximizes the capture rate $R$ and with a zero launch velocity ($\omega=\omega'$). During this
phase of duration $t_1$, the number of trapped atoms varies according to
$N(t)=(R/\gamma)\,(1-e^{-\gamma t})$, where $\gamma^{-1}$ is the mean escape time. One then switches
the detuning of the trapping lasers to a much larger value, with $\omega \neq \omega'$ set to provide
the required velocity $\bar v$. This launching phase must last a time $t_{2}$ so that all trapped
atoms can leave the injecting MOT and reach the magnetic guide. For a trap of length $L$ and a
distance $D$ between the trap exit and the guide, one has $t_{2}=(L+D)/\bar v$. The optimization of
$t_1$ depends on the values of $\gamma$ and the launching time $t_2$. For a low Rb vapor pressure
(small $R$ and small $\gamma$), the largest flux corresponds to
\begin{equation}
\gamma t_2 \ll 1 \quad : \qquad \Phi \simeq R \qquad \mbox{for} \qquad t_1=\sqrt{2t_2/\gamma}\ .
\label{optimumflux}
\end{equation}
In this case, the flux $\Phi$ in the guide is equal to the capture rate of the injecting MOT and the
operation in pulsed mode does not lead to a loss in efficiency. If we increase the Rb vapor pressure
in the cell and thus the rate $R$ so that $\gamma \sim t_2^{-1}$, the optimum operation of the pulsed
mode corresponds to
\begin{equation}
\gamma t_2 \sim 1 \quad : \qquad \Phi \sim 0.3\,R \quad \mbox{for} \quad t_1 \sim t_2 \sim
\gamma^{-1} \ . \label{optimumflux2}
\end{equation}

This pulsed mode  also makes the extraction of atoms from the injecting MOT much easier. Indeed, the
optical pumping of the atoms to the hyperfine level $F=1$ is not total, because of the stray light
present at the repumping frequency $|5S_{1/2},F=1\rangle\rightarrow |5P_{3/2},F=2\rangle$. Therefore,
after leaving the injecting MOT, the atoms may still be deflected by the residual radiation pressure
force resulting from an imbalance between the intensities of the various laser beams. This spurious
effect is strongly decreased if the detuning of these lasers is increased to a value $\delta \sim
-7\,\Gamma$ while the atoms travel in the ``dangerous" region.

 Finally, we note that, although the atomic beam is pulsed at
the entrance of the guide, the pulses broaden as they propagate because of the longitudinal velocity
dispersion $\Delta v$. This entails that a quasi-continuous beam is obtained after a distance $\sim L
\bar v/\Delta v$ if one chooses $t_1\sim t_2$.

\subsection{ Use of an auxiliary MOT}

\begin{figure}
\begin{center}
\includegraphics[width=6cm]{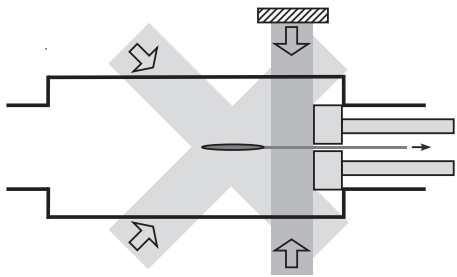}
\caption{\label{auxMOT}
 Use of an auxiliary 2D MOT to transfer the atoms from the injecting MOT to the magnetic guide.}
\end{center}
\end{figure}

Some of the difficulties associated with the free-flight transfer can also be circumvented if we
place an auxiliary guiding trap in the region $-25$~mm $<z<0$. This trap is a pure 2D MOT, whose
beams are orthogonal to the $z$ axis and illuminate points up to 4~mm from the entrance of the
magnetic guide (see fig.~\ref{auxMOT}). The transverse extension of the atomic beam at the entrance
of the magnetic guide is then drastically reduced. As for the free flight transfer, we place a
depumping laser at the entrance of the guide. The atoms are then trapped in the magnetic sublevel
$F=1,m=-1$.

The disadvantage of this auxiliary trap concerns the broadening of the longitudinal velocity
distribution, due to the random recoils associated with the spontaneous emission of photons as the
atoms cross the auxiliary trap. The longitudinal heating remains relatively weak for high atomic
velocities and large detunings $\delta_2$ of the auxiliary MOT. An upper limit for $|\delta_2|$ is
given only by the condition that the radiative force of the auxiliary MOT has to overcome the
spurious radiation pressure imbalance which exists in the wings of the injecting MOT's laser beams,
as discussed above.

\subsection{Experimental results}

 We now present the experimental results that we have obtained
concerning the characteristics of the guided atomic beam, in terms of temperature and flux. All these
measurements have been performed using the setup A, i.e. with a vapor-loaded injecting MOT.

\subsubsection{ Longitudinal velocity distribution}

The measurement of the longitudinal velocity distribution is quite straightforward if we operate the
injecting MOT in pulsed mode, since we can derive the longitudinal temperature directly from the
temporal width of the absorption signal when a given atom pulse passes through the probe beam. This
beam is located downstream, at a distance of either 40~cm or 200~cm from the entrance of the magnetic
guide.

 When operating in pulsed mode, we measure a longitudinal
temperature in the range 50-100~$\mu$K. The lowest temperatures are obtained for large velocities
($\bar v=2.6$~m/s). This effect is probably due to the heating of the atoms by the stray light of the
various trapping beams, when they travel over the distance $D$ between the injecting MOT and the
guide. This heating is larger if the atoms spend more time in this region, \emph{i.e.} if they are
slow. The presence of the auxiliary MOT causes some additional heating of the atomic beam,
proportional to the time spent by the atoms in the laser beams. For $\delta_2=-6.5\;\Gamma$, the
increase of the longitudinal temperature of the beam is at most $30\,$\%, while, as we shall see, the
transverse temperature decreases by an order of magnitude, thanks to the auxiliary MOT.

When the injecting MOT is operated in continuous mode, we found much larger longitudinal
temperatures, in the range of 0.5--1~mK. We think that this is due to the acceleration and heating of
the atoms when they travel between the injecting MOT and the guide, the heating being much more
dramatic than in the pulsed mode, since the frequencies of the trapping beams are closer to
resonance.

\subsubsection{ Transverse temperature}

 We have measured the transverse temperature in the guide  in
two ways, which give consistent results. First,  we have scanned transversally the position of the
probe laser beam and  reconstructed the transverse profile of the atomic beam. Alternatively, we have
used radio-frequency evaporation to selectively remove a fraction of the atomic distribution. From
the variation of the fraction of remaining atoms as a function of the radio-frequency, we infer the
transverse temperature of the atomic beam (see Appendix).

 Using the auxiliary MOT, we find a temperature ranging from
100 $\mu$K to $200\;\mu$K.  We have obtained similar results for the continuous and pulsed regimes.
The variation of the transverse temperature with the magnetic gradient $b'$ and the longitudinal
velocity is in good agreement with the estimate (\ref{temperature}), where $D$ should be replaced by
the distance $D'\sim 4$~mm between the exit of the auxiliary MOT and the entrance of the magnetic
guide.

 In absence of the auxiliary MOT, we have measured much larger
transverse temperatures inside the guide, in the range of 1-2~mK.
This is a consequence of the increase of the beam's transverse
size at the magnetic guide entrance, after free propagation over
the distance $D$. The variation of the transverse temperature with
$b'$ and $\bar v$ is also in good agreement with the prediction
(\ref{temperature}).

\subsubsection{ Flux}

The flux of atoms in the magnetic guide is not significantly modified by the presence of the
auxiliary MOT. Operating in continuous mode at a pressure $P_{87}=10^{-8}$~mbar, we find that this
flux varies between $1.5\times 10^8$ and $3\times 10^8$~atoms/s, when $\bar v$ varies between
$1.5$~m/s and $3$~m/s. This corresponds to a transfer efficiency between 15\% and 30\%. Similar
values are achieved in pulsed mode if we optimize the loading time $t_1$ according to
(\ref{optimumflux}).

For velocities smaller than 1.5~m/s, we did not find that a significant fraction of the atoms emitted
by the injecting MOT could be transferred to the magnetic guide. We attribute this fact to the
depletion of the slow atomic beam by collisions with atoms from the rubidium vapor in the cell. The
slow atoms have to travel over a distance $\sim 10$~cm through the cell and the tube ensuring the
differential pumping, before they arrive in the ultra-high vacuum region of the magnetic guide. For
$P_{87}=10^{-8}$~mbar (i.e. a total rubidium pressure $4\times 10^{-8}$~mbar) and $\bar v=1.5$~m/s,
we estimate the atomic flux to be reduced by 40~\% over this distance. This limitation should be
overcome with our new apparatus (setup B) as discussed at the end of \S~\ref{sec:injector}.

\subsection{ Miniature guide}

At low atomic velocities, the auxiliary MOT does not provide efficient coupling of atoms into the
guide due to strong recoil heating in the longitudinal direction. For this reason, we have developed
a miniature magnetic guide (MMG) on a conical supporting structure. The shape has been chosen in
order to keep a good optical access for the injecting MOT's beams. The MMG is added at the entrance
of the magnetic guide as depicted in fig.~\ref{setup}(b).

The MMG generates a two-dimensional magnetic quadrupole field by means of four elliptical
Kapton-coated copper wires with a 1 mm$^2$ cross section. The wires are placed at coordinates $x=\pm
b, y=\pm b$, with $b=1.3$ mm. They provide a transverse confinement over a distance of 40 mm up to a
point at which the confining force of the long magnetic guide has risen to its asymptotic value. The
inner supporting structure of the wires is a tube with a diameter $d=3$ mm, which also ensures
differential pumping between the injecting MOT's chamber and the section of magnetic guide.

 The recombination of the current between the wires of the MMG is
done in a way as to minimize the free flight distance. Consequently, the transverse free expansion
remains small. The longitudinal temperature is not affected in this case since no spontaneous
emission processes occur. The wires emerge from the supporting structure before being folded back
outwards and attached to the conical structure. The currents recombine in the metallic supporting
structure to which the non-isolated wire ends are screwed (see fig.~\ref{setup}(b)). In this way the
laser beams of the injecting MOT are not interrupted by the MMG's wires. The MMG allows to reduce the
distance over which the atoms expand freely to less than 3 mm. The magnetic gradient is typically 1
T/m (0.22 T m$^{-1}$A$^{-1}$). The current running in the MMG has to be chosen in the same sense as
the one in the guide. The wires of the MMG are kept cooled by thermal contact with the water-cooled
magnetic guide. With a current of 5 A running through the wires of the MMG, we find an increase of
resistivity that corresponds to a temperature rise of 5 K assuming an homogeneous temperature change
of the wires.

In order to measure the mean velocity of the atoms along the longitudinal axis, we use as before a
time-of-flight technique. We operate the injecting MOT in a pulsed mode. At time $t=0$, we launch a
bunch of atoms which propagates in the magnetic guide, and we monitor the time-dependent absorption
signal of a probe beam located further downstream. The MMG allows for the production of a very low
velocity beam in the magnetic guide. We have recently observed atoms in the magnetic guide with a
mean velocity of the order of 70 cm/s. For the production of such a low-velocity atomic beam, two
requirements have to be met. The injecting MOT's beam profiles have to be balanced over the whole
free-flight distance (otherwise the atoms are accelerated while passing through the wings of the
laser beams). Furthermore, care has to be taken that the atoms are not pumped out of the
$|F=1,m_f=-1\rangle$ level by stray light while propagating in the magnetic guide. For that purpose,
we detune the frequency of the repumping light by several line widths. Actually, we do not detect any
atoms in the guide when the repumping laser is on resonance.

In conclusion, we have reported the continuous and pulsed loading of a slow and cold atomic beam into
a magnetic guide. In order to optimize the transfer into the guide, we have investigated various
coupling schemes. The first one is based on an auxiliary two-dimensional MOT that confines the atoms
transversally before they enter the guide. However, for low atomic velocities, the atoms are strongly
heated in the longitudinal direction. This limitation does not occur in another coupling method which
relies on magnetic confinement in the transfer zone. For this purpose, we have constructed a
miniature magnetic guide located between the injecting MOT and the long magnetic guide. The latter
scheme allows to inject atoms with velocities down to 70 cm/s into the guide, a result which we hope
to further improve in future experiments.

\section*{acknowledgements}
This work was supported by the Bureau National de la M\'etrologie, the D\'el\'egation G\'en\'erale de
l'Armement, the Centre National de la Recherche Scientifique and the R\'egion Ile de France. C. F.
Roos acknowledges support from the European Union (contract HPMFCT-2000-00478).

\appendix
\section{Temperature measurements by evaporation in the collisionless
regime}
\label{A}

The appendix is devoted to the  determination of the transverse temperature from the fraction of
remaining atoms when a radio frequency wave is applied. We restrict the analysis to the case of a
collisionless gas.  The atoms propagate freely along the guide axis ($z$) and they are confined in
the transverse $xy$ plane by an isotropic harmonic potential with frequency $\omega$. The transverse
and longitudinal motions are decoupled from each other and the total Hamiltonian then reads:
\[
H({\bf r},{\bf p})=H_\bot({\bf r}_\bot,{\bf p}_\bot)
+\frac{p_z^2}{2m}\ ,
\]
where the Hamiltonian for the transverse motion is:
\[
H_\bot({\bf r}_\bot,{\bf
p}_\bot)=\frac{p^2}{2m}+\frac{1}{2}m\omega^2 r^2
\]
with $p^2=p_x^2+p_y^2$, $r^2=x^2+y^2$. Another dynamical quantity
of interest for the study of the transverse motion is the
$z$-component of the angular momentum
\[
{\cal L}({\bf r}_\bot,{\bf p}_\bot)=xp_y-yp_x\ .
\]

 The trajectory of an atom is characterized by three constants
of motion: the velocity  $v$ along the guide axis, the total  (kinetic+potential) transverse energy
$E=H_\bot({\bf r}_\bot,{\bf p}_\bot)$, and the  $z$-component of the angular momentum $L={\cal
L}({\bf r}_\bot,{\bf p}_\bot)$. One readily establishes a relation between the two quantities $E$ and
$L$ characterizing the transverse motion: $E\ge |L|\omega$. For a given $(E,L)$, the trajectory is
confined in the region of space $r_{\rm min} \le r \le r_{\rm max}$ with:
 \[
r^2_{\rm min}=\frac{E-\sqrt{E^2-L^2\omega^2}}{m\omega^2}\ \ \ \
r^2_{\rm max}=\frac{E+\sqrt{E^2-L^2\omega^2}}{m\omega^2}.
 \]

The evaporation is performed by inducing spin flips with a radiofrequency field of frequency $\nu$ on
a surface defined by: $\mu |B({\bf r}_{\rm evap})|=h\nu$. Note that it does not mean {\it a priori}
that atoms with an energy higher than $h\nu$ are evaporated. It depends actually on the total energy
of the particle and on its angular momentum. Evaporation occurs only if $r_{\rm min}\le r_{\rm evap}
\le r_{\rm max}$, which means that the trajectory goes through the  surface of evaporation. In the
following, we note $E_0=m\omega^2r^2_{\rm evap}$.

In order to determine the fraction $f$ of atoms  which is not affected by the evaporation, one needs
to derive the density probability $P(E,L)$ for a given atom to have an energy $E$ and  an angular
momentum $L$. By definition,
 \begin{equation}
 P(E,L) \propto \int d^2r\; d^2p\; e^{-\beta H_\bot}\;
 \delta(L-{\cal L})\;\delta (E-H_\bot)\ ,
 \end{equation}
where $\beta=1/(k_BT)$.   One readily obtains \cite{rq1}:
 \[
 P(E,L)\propto e^{-\beta E}\Theta(E-|L|\omega)\; ,
 \]
where $\Theta$ is the Heaviside step function. We distinguish two categories of atoms that are not
affected by the radio-frequency wave:
\begin{itemize}
\item
atoms such that $r_{\rm min}>r_{\rm evap}$, the corresponding range of values for $E$ and $L$ is
denoted $\mbox{$\mathcal{D}$}_1$;

\item
atoms such that $r_{\rm max} < r_{\rm evap}$, the corresponding range of values for $E$ and $L$ is
denoted $\mbox{$\mathcal{D}$}_2$.
\end{itemize}

 The fraction $f$ of non-evaporated atoms is obtained after a
simple calculation and it reads:
 \begin{equation}
 f=\int\limits_{\mbox{$\mathcal{D}$}_1 \mbox{$\cup$}
 \mbox{$\mathcal{D}$}_2}P(E,L)\;dE\; dL=1-
 \sqrt{\frac{\pi\beta E_0}{2}}e^{-\beta
 E_0/2}.
 \label{fevap}
 \end{equation}
The temperature is obtained by fitting  the experimental data to the function (A3). This fraction has
a minimum for $r_{\rm evap}=(\beta m\omega^2)^{-1/2}$ equal to $f_{\rm min} \simeq 0.2398...$.
Without  elastic collisions, it is impossible to evaporate all atoms. On the other hand, since
collisions allow the redistribution of angular momentum and energy, the observation of a fraction of
non-evaporated atoms lower than $f_{\rm min}$ is a signature of the presence of elastic collisions in
the sample when crossing the region where evaporation is applied. We emphasize that this treatment
 is valid only if the magnetic potential is close to a harmonic potential: $k_BT \ll \mu B_0$,
where $T$ is the temperature, $\mu$ the magnetic moment and $B_0$ the longitudinal bias field applied
in the evaporation region. When this criterium is not fulfilled, we have to use a numerical model
based on a Monte-Carlo sampling of the atomic distribution rather than the analytical expression
(\ref{fevap}).

$^*$ Unit\'e de Recherche de l'Ecole normale sup\'erieure et de l'Universit\'e Pierre et Marie Curie,
associ\'ee au CNRS.

\end{document}